\documentclass[11pt,a4paper]{article}
\usepackage{moriond,epsfig}

\usepackage{xspace}

\usepackage{caption}
\usepackage{subfig}

\usepackage{ifthen}
\newboolean{uprightparticles}
\setboolean{uprightparticles}{true}
\usepackage{amssymb}
\usepackage{amsfonts}
\usepackage{upgreek}

\def\lhcb  {LHCb\xspace}
\def\LHCb  {\lhcb}
\def\cern  {CERN\xspace}
\def\lhc   {LHC\xspace}

\def\hfag  {HFAG\xspace}
\def\HFAG  {\hfag}

\ifthenelse{\boolean{uprightparticles}}%
{
 \def\Ppi      {\ensuremath{\uppi}\xspace}

 \def\PB      {\ensuremath{\mathrm{B}}\xspace}                 
 \def\PD      {\ensuremath{\mathrm{D}}\xspace}                 
 \def\PK      {\ensuremath{\mathrm{K}}\xspace}

 \def\Pp      {\ensuremath{\mathrm{p}}\xspace}                 
}
{
 \def\Ppi      {\ensuremath{\pi}\xspace}

 \def\PB      {\ensuremath{B}\xspace}                 
 \def\PD      {\ensuremath{D}\xspace}                 
 \def\PK      {\ensuremath{K}\xspace}

 \def\Pp      {\ensuremath{p}\xspace}                 
}

\def\pion  {\ensuremath{\Ppi}\xspace}
\def\pip   {\ensuremath{\pion^+}\xspace}
\def\pim   {\ensuremath{\pion^-}\xspace}

\newcommand{\pislow}{\ensuremath{\pion_{\mathrm{slow}}}\xspace}

\def\kaon  {\ensuremath{\PK}\xspace}
\def\Kp    {\ensuremath{\kaon^+}\xspace}
\def\Km    {\ensuremath{\kaon^-}\xspace}

\def\Dbar    {\kern 0.2em\overline{\kern -0.2em \PD}{}\xspace}
\def\D       {\ensuremath{\PD}\xspace}
\def\Db      {\ensuremath{\Dbar}\xspace}
\def\Dz      {\ensuremath{\D^0}\xspace}
\def\Dzb     {\ensuremath{\Dbar^0}\xspace}
\def\DzDzb   {\ensuremath{\Dz {\kern -0.16em \Dzb}}\xspace}
\def\Dp      {\ensuremath{\D^+}\xspace}
\def\Dm      {\ensuremath{\D^-}\xspace}

\def\DpDm    {\ensuremath{\Dp {\kern -0.16em \Dm}}\xspace}
\def\Dstar   {\ensuremath{\D^*}\xspace}

\def\Dstarp  {\ensuremath{\D^{*+}}\xspace}
\def\Dstarm  {\ensuremath{\D^{*-}}\xspace}

\def\B       {\ensuremath{\PB}\xspace}
\def\Bbar    {\kern 0.18em\overline{\kern -0.18em \PB}{}\xspace}

\def\proton      {\ensuremath{\Pp}\xspace}

\def\CP                {\ensuremath{C\!P}\xspace}

\def\CPV               {\ensuremath{C\!PV}\xspace}

\newcommand{\ARaw}{\ensuremath{{\cal A}_{\mathrm{Raw}}}\xspace}
\newcommand{\ARawstar}{\ensuremath{\ARaw^{\ast}}\xspace}
\newcommand{\AD}{\ensuremath{{\cal A}_{\mathrm{D}}}\xspace}
\newcommand{\AP}{\ensuremath{{\cal A}_{\mathrm{P}}}\xspace}

\newcommand{\Delm}{\mbox{$\Delta m $}\xspace}
\newcommand{\ACP}{\ensuremath{{\cal A}_{\CP}}\xspace}

\def\ycp        {\ensuremath{y_{\CP}}\xspace}
\def\yCP        {\ycp}
\def\agamma     {\ensuremath{A_{\Gamma}}\xspace}
\def\AGamma     {\agamma}

\newcommand{\tev}{\ensuremath{\mathrm{\,Te\kern -0.1em V}}\xspace}
\newcommand{\gev}{\ensuremath{\mathrm{\,Ge\kern -0.1em V}}\xspace}
\newcommand{\mev}{\ensuremath{\mathrm{\,Me\kern -0.1em V}}\xspace}
\newcommand{\kev}{\ensuremath{\mathrm{\,ke\kern -0.1em V}}\xspace}
\newcommand{\ev}{\ensuremath{\mathrm{\,e\kern -0.1em V}}\xspace}
\newcommand{\gevc}{\ensuremath{{\mathrm{\,Ge\kern -0.1em V\!/}c}}\xspace}
\newcommand{\mevc}{\ensuremath{{\mathrm{\,Me\kern -0.1em V\!/}c}}\xspace}
\newcommand{\gevcc}{\ensuremath{{\mathrm{\,Ge\kern -0.1em V\!/}c^2}}\xspace}
\newcommand{\gevgevcccc}{\ensuremath{{\mathrm{\,Ge\kern -0.1em V^2\!/}c^4}}\xspace}
\newcommand{\mevcc}{\ensuremath{{\mathrm{\,Me\kern -0.1em V\!/}c^2}}\xspace}

\def\mbarn{\ensuremath{\rm \,mb}\xspace}

\def\invpb {\ensuremath{\mbox{\,pb}^{-1}}\xspace}

\def\invfb   {\ensuremath{\mbox{\,fb}^{-1}}\xspace}

\newcommand{\stat}{\ensuremath{\mathrm{(stat)}}\xspace}
\newcommand{\syst}{\ensuremath{\mathrm{(syst)}}\xspace}

\def\lumiyr {\ensuremath{37.7\,\invpb}\xspace}
\def\lumiacp {\ensuremath{37\,\invpb}\xspace}

\usepackage{url}

\def\be{\begin{equation}}
\def\ee{\end{equation}}
\def\bea{\begin{eqnarray}}
\def\eea{\end{eqnarray}}

\begin{document}
\vspace*{4cm}
\title{CHARM PHYSICS RESULTS AND PROSPECTS WITH \LHCb}

\author{P.M. SPRADLIN\,\footnote{On behalf of the \LHCb Collaboration}}

\address{Department of Physics and Astronomy, University of Glasgow, Kelvin Building, University Avenue,\\
Glasgow G12 8QQ, United Kingdom}

\maketitle\abstracts{
  Precision measurements in charm physics offer a window into a unique 
  sector of potential New Physics interactions.
  \LHCb is poised to become a world leading experiment for charm physics,
  recording enormous samples with a detector tailored for flavor physics.
  This article presents recent charm production, \CPV, and mixing studies from
  \LHCb, including \LHCb's first charm \CP asymmetry measurement with \lumiacp
  of data collected in 2010.
  Significant updates to the material presented at the 2011 Rencontres de
  Moriond QCD and High Energy Interactions are included.
}

\section{Charm production at the \LHCb experiment}
\label{sec:LHCb}

  \LHCb\cite{Alves:2008zz}, the dedicated flavor experiment at \cern's Large
  Hadron Collider (\lhc), is the only \lhc experiment with a broad charm
  physics program including measurements of charm \CP violation (\CPV) and
  $\Dz$-$\Dzb$ mixing.
  The cross-section to produce charm hadrons into the \LHCb acceptance
  in the LHC's $\sqrt{s} = 7\,\tev$ proton-proton collisions is
  \mbox{$1.23 \pm 0.19\,\mbarn$}, creating a huge potential data
  set\@.\cite{LHCb-CONF-2010-013}
  The \LHCb trigger system has a flexible design that includes
  charm triggers so that this prolific production can be exploited.

  \LHCb recorded a total integrated luminosity of \lumiyr in 2010.
  The charm samples collected in 2010 are already large enough for \LHCb to
  be competitive in several measurements.
  With an expectation of more than $1\,\invfb$, the 2011-12 run will yield
  even larger samples.

  Because the \lhc collides protons, there may be asymmetries in the production
  of charm and anti-charm hadrons.
  \LHCb has measured the production asymmetry of \Dz/\Dzb using \lumiacp of
  2010 data\@.\cite{LHCb-CONF-2011-023}
  The analysis uses both untagged samples of reconstructed \Dz decays
  and tagged samples that are reconstructed as the product of a
  $\Dstarp \to \Dz \pislow^+$ decay.
  In the tagged sample, the initial flavor of the \D is
  identified (tagged) as \Dz or \Dzb by the charge of the tagging slow
  pion, $\pislow^\pm$.
  In both samples, \Dz is reconstructed in the final states
  $\Km\pip$, $\Km\Kp$, and $\pim\pip$.
  For a final state $f$, the raw observed untagged asymmetry, $\ARaw(f)$, and
  the raw observed \Dstar-tagged asymmetry, $\ARawstar(f)$, can be factored
  into components:
  \begin{eqnarray}
    \ARaw(f) & \equiv & \frac{N(\Dz \to f) - N(\Dzb \to \bar{f})}{N(\Dz \to f) + N(\Dzb \to \bar{f})}   =  \ACP(f) + \AD(f) + \AP(\Dz),\label{eq:cpv:comp:untag} \\
    \ARawstar(f) & \equiv & \frac{N(\Dstarp \to \Dz(f)\pislow^+) - N(\Dstarm \to \Dzb(\bar{f})\pislow^-)}{N(\Dstarp \to \Dz(f)\pislow^+) + N(\Dstarm \to \Dzb(\bar{f})\pislow^-)}\nonumber \\
        & = & \ACP(f) + \AD(f) + \AD(\pislow) + \AP(\Dstarp),\label{eq:cpv:comp:tag}
  \end{eqnarray}
  where the $N(\mbox{decay})$ are the numbers of reconstructed decays,
  $\ACP(f)$ is the \CP asymmetry of the \Dz decay (further studied in
  Section~\ref{sec:cpv}), $\AD(f)$ and
  $\AD(\pislow)$ are the detection asymmetries of $f$ and $\pislow^{\pm}$, and
  $\AP(\Dz)$ and $\AP(\Dstarp)$ are the production asymmetries.
  For the self-conjugate final states $\Km\Kp$ and $\pim\pip$,
  $\AD(\Km\Kp) = \AD(\pim\pip) = 0$.
  Therefore, the remaining detection asymmetries can be canceled by
  considering combinations of raw asymmetries,
  \begin{eqnarray}
    \ARaw(\Km\pip) - \ARawstar(\Km\pip) + \ARawstar(\Km\Kp) & = & \AP(\Dz) + \ACP(\Km\Kp),\label{eq:prod:kk} \\
    \ARaw(\Km\pip) - \ARawstar(\Km\pip) + \ARawstar(\pim\pip) & = & \AP(\Dz) + \ACP(\pim\pip).\label{eq:prod:pp}
  \end{eqnarray}
  Using the \HFAG world averages of $\ACP(\Km\Kp)$ and
  $\ACP(\pim\pip)$\cite{Asner:2010qjmod} and a Bayesian minimizer to optimally
  solve this over-constrained system for $\AP(\Dz)$, we measure a mean value of
  $\AP(\Dz) = \left[-1.08 \pm 0.32\,\stat \pm 0.12\,\syst \right]\%$ in
  \LHCb's acceptance.

\section{Time-integrated \CPV in \D mesons}
\label{sec:cpv}

  \LHCb is searching for evidence of new sources of \CP asymmetry in the
  time-integrated decay rates of \D mesons.
  The time-integrated \CP asymmetry, $\ACP(f)$, is conventionally defined as
  \begin{equation}
    \ACP(f) = \frac{\Gamma(\D \to f) - \Gamma(\Db \to \bar{f})}{\Gamma(\D \to f) + \Gamma(\Db \to \bar{f})}
    \label{eq:cpv:acp}
  \end{equation}
  for a given final state $f$.
  For \Dz decays, $\ACP$ may have contributions from both indirect and direct
  \CPV\@.
  In the Standard Model, \CPV in the charm system is highly suppressed.
  Indirect \CPV is negligibly small and should be common for all decay modes.
  Direct \CPV is expected to be $\mathcal{O}(10^{-3})$ or less and to vary
  among decay modes\@.\cite{Bianco:2003vb}
  In \CPV searches in singly Cabibbo suppressed decays,
  such as $\Dz \to \Km \Kp$, participation of well-motivated new physics
  (NP) particles in the interfering penguin amplitude
  could enhance direct \CPV up to $\mathcal{O}(10^{-2})$\@.\cite{Grossman:2006jg}

  \LHCb recently presented its first time-integrated \CPV
  measurement with decays $\Dz \rightarrow \Km \Kp$ and
  $\Dz \rightarrow \pim \pip$\@.\cite{LHCb-CONF-2011-023}
  The analysis uses the tagged samples of $\Dstarp \rightarrow \Dz \pislow^+$
  decays also used in the measurement of $\AP(\Dz)$ (Section~\ref{sec:LHCb}).
  Using Equation~\ref{eq:cpv:comp:tag}, the difference in $\ACP(f)$ for
  $f = \Km\Kp$ and $\pim\pip$ can be measured precisely with the
  production and detection asymmetries canceling exactly:
  \begin{eqnarray}
    \Delta\ACP   & \equiv & \ACP(\Km\Kp) - \ACP(\pim\pip),\label{eq:cpv:delacp:def} \\
                 & = & \ARawstar(\Km\Kp) - \ARawstar(\pim\pip).\label{eq:cpv:delacp:raw}
  \end{eqnarray}
  In \lumiacp of \LHCb 2010 data, we measure
  $\Delta\ACP = \left[-0.28 \pm 0.70\,\stat \pm 0.25\,\syst \right]\%$,
  consistent with zero.
  This result is approaching the sensitivity of \CPV measurements performed by
  the \B-factories in these decay modes,\cite{Aubert:2007if,Staric:2008rx}
  but not yet at the level of CDF's recent measurement\@.\cite{CDF-10296}
  Due to differential proper-time acceptance between the $\Km\Kp$ and
  $\pim\pip$ samples, the measured value of $\Delta\ACP$ includes a residual
  $10\%$ of the mode-independent indirect \CP asymmetry.
  No limiting systematic bias has been identified in the method, so
  future iterations of the measurement with the much larger data set
  anticipated for 2011-2012 will be significantly more precise.

\section{Time-dependent \CPV and mixing measurements in \Dz}
\label{sec:mix}

  The conventional parameterization of charm mixing is fully explained
  elsewhere\@.\cite{Nakamura:2010zzi:D0mix}
  Briefly, the mass eigenstates of the neutral \D system $\D_1$ and $\D_2$
  are expressed as normalized superpositions of the flavor eigenstates \Dz
  and \Dzb, $\D_{1,2} = p \Dz \pm q \Dzb$, where $p$ and $q$ are complex
  scalars, $|p|^2 + |q|^2 = 1$.
  The relative argument of $q$ and $p$ is conventionally chosen equal to
  the phase that parameterizes \CPV in the interference between mixing and
  direct decays, $\arg\frac{q}{p} = \phi$.
  \CP is violated in the mixing if $|\frac{q}{p}| \ne 1$ and in
  the interference between mixing and decay if $\phi \ne 0$.
  Letting $m_{1,2}$ and $\Gamma_{1,2}$ represent respectively the masses and
  widths of $\D_{1,2}$, mixing is parameterized by
  the mass difference $x \equiv \frac{m_1 - m_2}{\Gamma}$
  and the width difference $y \equiv \frac{\Gamma_1 - \Gamma_2}{2 \Gamma}$
  where $\Gamma \equiv \frac{1}{2}\left(\Gamma_1 + \Gamma_2 \right)$.

  \LHCb is working towards its first measurements of \CPV and mixing in
  \Dz-\Dzb with lifetime ratios of \mbox{$\Dz \rightarrow \Km \pip$} and
  \mbox{$\Dz \rightarrow \Km \Kp$} decays.
  The lifetime of decays to the \CP-even eigenstate $\Km\Kp$, $\tau(\Km\Kp)$,
  is related to the lifetime of the flavor-specific final state $\Km\pip$,
  $\tau(\Km\pip)$, by the mixing parameters:
  \begin{equation}
    \yCP \equiv \frac{\tau(\Km\pip)}{\tau(\Km\Kp)} - 1 = y \cos\phi 
        - \frac{1}{2}\left(\left|\frac{q}{p}\right| - \left|\frac{p}{q}\right|\right) x \sin\phi.
    \label{eq:mix:ycp}
  \end{equation}
  If \CP is conserved, $\yCP = y$.
  The asymmetry in the lifetimes of \Dz and \Dzb decays to the \CP eigenstate
  $\Km\Kp$ is related to the \CPV and mixing parameters by
  \begin{equation}
    \AGamma \equiv \frac{\tau(\Dzb \to \Km\Kp) - \tau(\Dz \to \Km\Kp)}{\tau(\Dzb \to \Km\Kp) + \tau(\Dz \to \Km\Kp)}
        = \frac{1}{2}\left(\left|\frac{q}{p}\right| - \left|\frac{p}{q}\right|\right) y \cos\phi - x \sin\phi.
    \label{eq:mix:Agamma}
  \end{equation}
  \Dstar-tagged candidates are used in the measurement of \AGamma, while \yCP
  can be measured with the larger untagged sample.

  \begin{figure}[hpb]

      \subfloat[$\Dstarp \to \Dz \pip$]{\label{fig:mix:deltam:Dz}\psfig{file=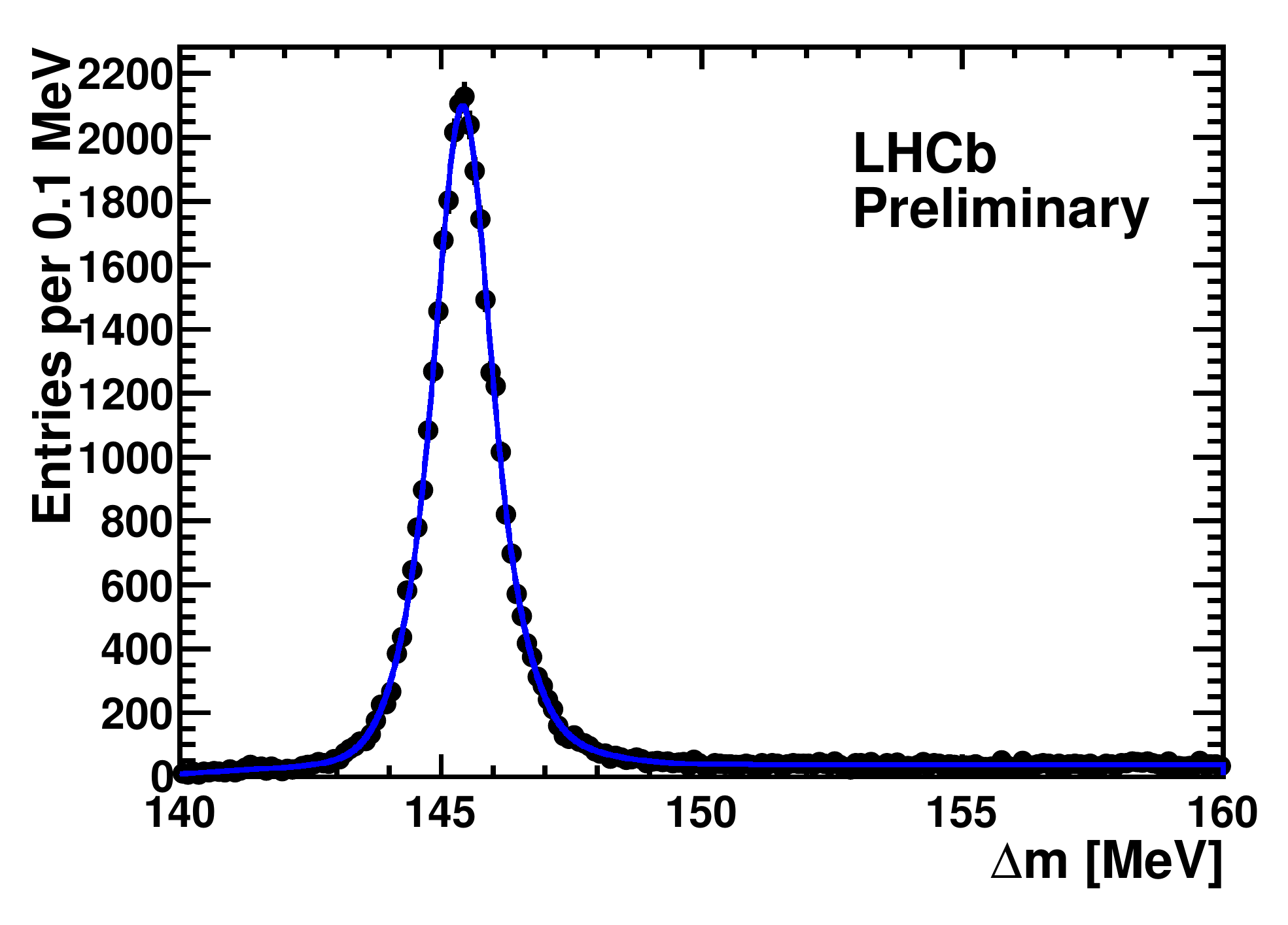,width=0.5\textwidth}}%
      \subfloat[$\Dstarm \to \Dzb \pim$]{\label{fig:mix:deltam:Dzb}\psfig{file=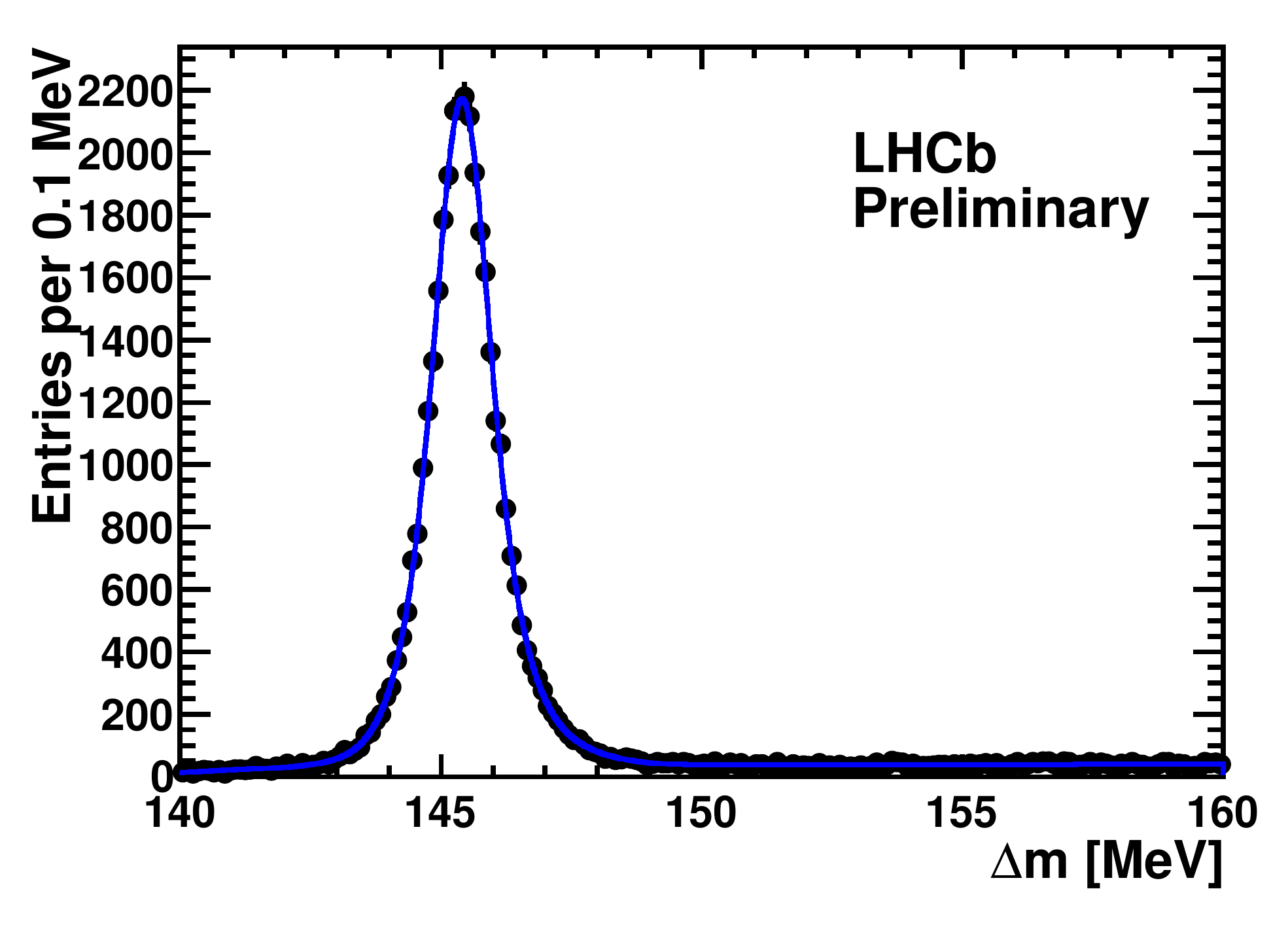,width=0.5\textwidth}}

    \caption{Distributions of the mass difference, $\Delta m$, between
      \Dz(\Dzb) candidates and their parent \Dstarp(\Dstarm) candidates for
      decays $\Dstarp \rightarrow \Dz \pip$,
      $\Dz \rightarrow \Km \pip$ (c.c.).
     \label{fig:mix:deltam}}
  \end{figure}

  In the 2010 run, we collected a sample of untagged
  \mbox{$\Dz \to \Km \Kp$} decays comparable in size to those of recent
  Belle and BaBar measurements\@.\cite{Staric:2007dt,Auber:2009ck}
  In 2011-2012, \LHCb expects to have the world's largest charm sample in this
  mode.
  The measurements of \yCP and \AGamma are currently blinded.
  As a test, the \AGamma analysis was applied to a subset of the
  2010 data in the right-sign (RS) control channel $\Dz \rightarrow \Km \pip$.
  Figure~\ref{fig:mix:deltam} shows the distributions of the differences
  \Delm between the masses of the reconstructed \Dz candidates and their
  parent \Dstarp candidates for the RS validation sample.
  The purity of the sample is better than $90\%$.

\begin{figure}[ht]

      \subfloat[\Dz]{\label{fig:mix:t:Dz}\psfig{file=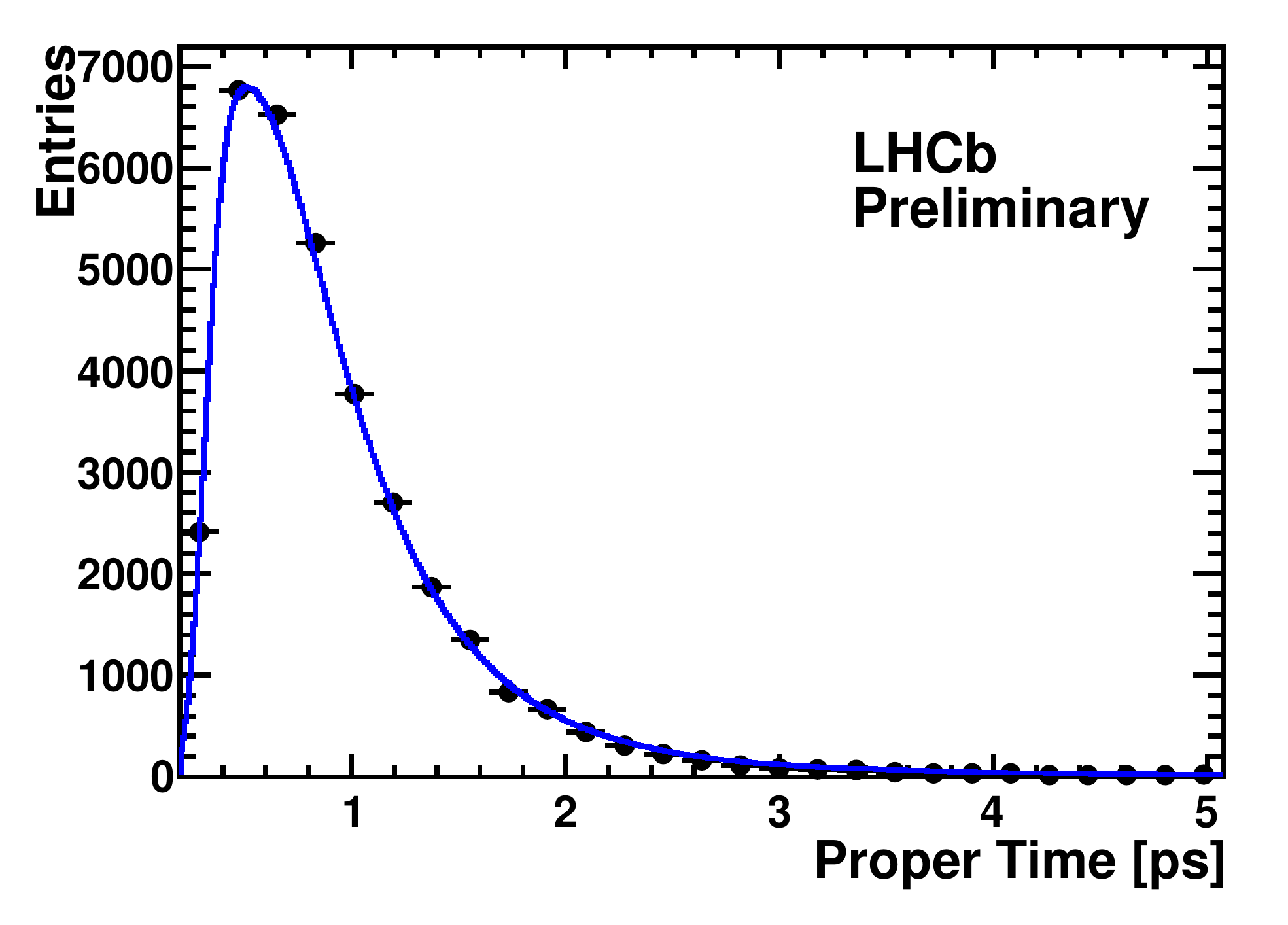,width=0.5\textwidth}}%
      \subfloat[\Dzb]{\label{fig:mix:t:Dzb}\psfig{file=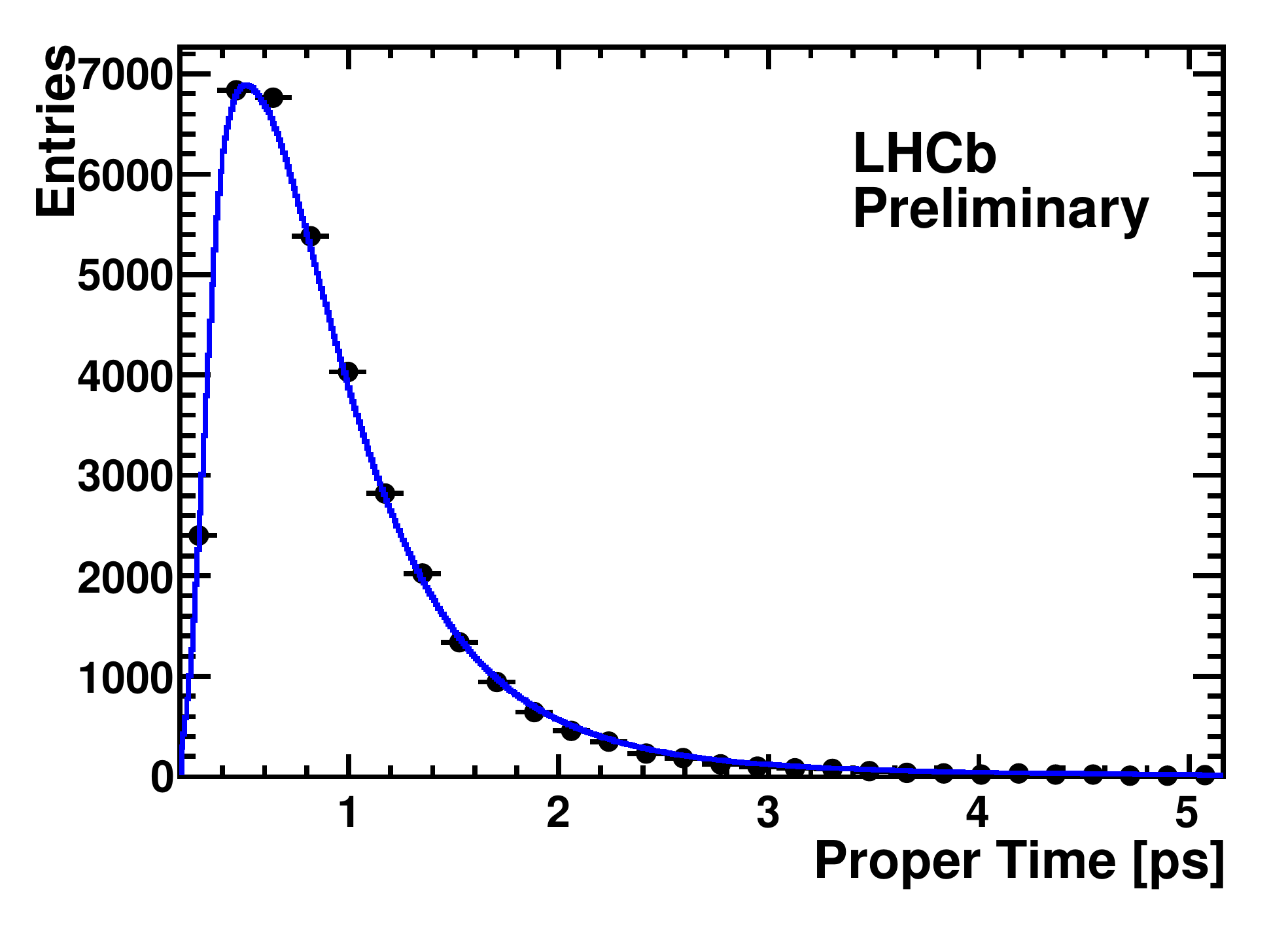,width=0.5\textwidth}}

    \caption{Distributions of the reconstructed proper time of \Dz(\Dzb)
      candidates for decays $\Dstarp \rightarrow \Dz \pip$,
      $\Dz \rightarrow \Km \pip$ (c.c.).
      The line on each plot is the result of a likelihood fit incorporating
      per-event acceptance distributions computed with the swimming method.
     \label{fig:mix:t}}
  \end{figure}

  Since the most powerful signal/background discriminants in hadronic
  collisions exploit the relatively long lifetime of \D mesons,
  the trigger and selection criteria introduce a proper-time acceptance for
  the reconstructed \Dz decays.
  Unbiased time-dependent measurements require careful treatment of the
  acceptance effects of these discriminants.
  \LHCb can precisely evaluate the proper-time acceptance on an event-by-event
  basis with the swimming method\@.\cite{Gersabeck:1217589,Aaltonen:2010ta}
  Statistical separation of \Dz mesons produced at the primary interaction
  vertex from those produced in the decays of $b$-hadrons is accomplished
  using the impact parameter (IP) $\chi^2$ of the \Dz.
  The event-by-event acceptance and the IP $\chi^2$ are incorporated into an
  unbinned multi-dimensional likelihood fit to measure the lifetimes.
  Figure~\ref{fig:mix:t} shows the proper-time distributions for the tagged
  RS validation sample.
  The lines on the plots are the fitted distributions from the unbinned
  multi-dimensional likelihood fit.

\section{Summary}
\label{sec:sum}

  \LHCb had a successful year of data taking in 2010, collecting \lumiyr of
  $\proton\proton$ collisions at $\sqrt{s} = 7\,\tev$.
  We observe an asymmetry in \Dz production of
  $\AP(\Dz) = [-1.08 \pm 0.32\,\stat \pm 0.12\,\syst]\%$, which is
  the first evidence for an asymmetry in heavy flavor production at the \lhc.
  In our first precision charm \CPV measurement with this
  data, the difference between the time-integrated \CP asymmetries of
  $\Dz \rightarrow \Km\Kp$ and $\Dz \rightarrow \pim\pip$ decays is measured
  to be $\Delta\ACP = [-0.28 \pm 0.70\,\stat \pm 0.25\,\syst]\%$.
  A broad program of charm physics is underway and
  further results in more channels are soon to follow.
  With the large data set expected in 2011-2012, \LHCb is poised to become a
  leader in charm physics.

\section*{References}

\providecommand{\href}[2]{#2}
\begingroup\raggedright\endgroup

\end{document}